\documentclass[12pt,amsmath,amssym,latexsymb]{article}
\usepackage{amssymb,latexsym}
\usepackage{longtable,tabularx}
\usepackage{makeidx}
\usepackage{graphicx}
\usepackage{psfrag}

\def\tilde{\widetilde}
\def \bfo {\begin {eqnarray*} }
\def \efo {\end {eqnarray*} }
\def \ba {\begin {eqnarray*} }
\def \ea {\end {eqnarray*} }
\def \beq {\begin {eqnarray}}
\def \eeq {\end {eqnarray}}

\def\picture #1 by #2 (#3){
\vsquare to #2{
            \hrule width #1 height 0pt depth 0pt
            \hfill
            \special{picture #3}}}

\def\scaledpicture #1 by #2 (#3 scaled #4){{
          \dimen0=#1 \dimen1=#2
          \divide\dimen0 by 1000 \multiply\dimen0 by #4
          \divide\dimen1 by 1000 \multiply\dimen1 by #4
          \picture \dimen0 by \dimen1 (#3 scaled #4)}}

\def \radius {R}

\newcommand{\R}{{\mathbb R}}

\def \H2s {H^{s+1}_0(\partial M\times [0,T/2])}

\def \det {\hbox{det}}

\def \e {\varepsilon}

\def \pa0 {\partial _0}

\def \p {\partial}

\def\e{\varepsilon}

\def\M{{M}}

\def\tilde{\widetilde}

\def \mbeq {\begin {eqnarray}}
\def \meeq {\end {eqnarray}}

\def \bfo {\begin {displaymath} }
\def \efo {\end {displaymath} }
\def \beq {\begin {eqnarray}}
\def \eeq {\end {eqnarray}}
\def \ba {\begin {eqnarray*}}
\def \ea {\end  {eqnarray*}}

\def \H2s {H^{s+1}_0(\partial \M\times [0,T/2])}

\def \det {\hbox{det}}

\def \e {\varepsilon}

\def \pa0 {\partial _0}

\def \p {\partial}

\def \tilde{\widetilde}

\newcommand{\newtekst}{}

\begin{document}

\title{
Effectiveness and improvement of cylindrical cloaking with the SHS
lining\\
}

\author{
Allan Greenleaf\\
Yaroslav Kurylev\\
Matti Lassas\\
Gunther Uhlmann}

\date{}

\maketitle

\begin{abstract}
We analyze, both analytically and numerically, the effectiveness of cloaking
 an infinite cylinder from observations by electromagnetic waves in
 three dimensions. We show that, as truncated approximations 
of the ideal permittivity and permeability {\newtekst tensors} 
tend towards the singular
 ideal cloaking fields, so that the anisotropy ratio tends to infinity, 
{\newtekst the $D$ and $B$ fields blow up near the cloaking surface.} 
Since the metamaterials used to implement
 cloaking are based on effective medium theory, the resulting large variation 
in $D$ and $B$  will pose a challenge to the suitability of the field averaged
 characterization of $\e$ and $\mu$. We also consider cloaking with and 
without the SHS (soft-and-hard surface) lining, shown in \cite{GKLU1} 
to be theoretically necessary for cloaking  in the cylindrical geometry. 
We demonstrate numerically that
cloaking is 
significantly improved by the SHS lining, with both the far field of the scattered 
wave  significantly reduced  and the blow up of $D$ and $B$  prevented.

\end{abstract}

\section{Introduction}

\subsection{Background and history}

There has recently been much activity concerning \emph{cloaking}, or rendering
objects invisible to detection by electromagnetic (EM) waves. For theoretical
descriptions  of EM material parameters of the general type
considered here, see
\cite{GLU1,GLU2,Le,PSS1,PSS2,GKLU1}; for numerical and experimental results,
see \cite{CPSSP,SMJCPSS,CCKS,CC,ZGNP}. Related results  concerning elastic waves are
in \cite{MBW,CS,M}. All of these papers treat cloaking  in the frequency domain,
using time harmonic waves of some fixed frequency $k\ge 0$; this is not
unreasonable, since the metamaterials used to implement these designs 
seem to be 
inherently prone to dispersion, for both practical and theoretical reasons
\cite{SF,Mo,PSS1}. See \cite{W} for a treatment of cloaking in the time domain.
One can also design electromagnetic  wormholes,  
which allow the passage of waves between possibly distant points 
while most of the wormhole remains invisible \cite{GKLU2,GKLU3}.

When physically constructing a cloaking (or wormhole) device, one is of
course not  able to exactly match the ideal description of the EM material
parameters (electric permittivity $\epsilon$ and magnetic
permeability $\mu$, for the
purposes of this paper). Any actual implementation will only realize a discrete
sampling of the values of $\e$ and $\mu$, and not be able
to assume the ideal  values
at points $x$ on the cloaking surface(s), where the tensors
$\e(x)$ or $\mu(x)$ have 0 or $\infty$ as eigenvalues.

\subsection{Approximate cloaking and linings}

The purpose of the current paper is twofold. First, we wish to explore the
degradation of
cloaking that occurs when the ideal material parameter fields are replaced with
approximations  obtained by limiting the \emph{anisotropy ratio},
$L$, as described
below.  This was  studied in two very interesting
recent papers.
Ruan, Yan, Neff and Qiu \cite{RYNQ} consider the effect on cloaking of truncation of the the
material parameter
fields, while Yan, Ruan and Qiu \cite{YRQ}, study the effect
of using the simplified material parameters employed in \cite{SMJCPSS,CCKS}. 
In \cite{RYNQ}, it is shown that cloaking of  passive objects, i.e., those 
with internal current $J=0$, holds in the limit as $L\to\infty$, but a slow rate of 
convergence of the fields is noted. The current paper reproves this and demonstrates the 
 blow up of the $B$ and $D$ fields at the cloaking surface as 
 $L\to \infty$.

Secondly, we consider the
effect of either including or not  including a physical lining to implement the
\emph{soft-and-hard surface} (SHS) boundary  condition, which is a boundary condition 
originally introduced in antenna design \cite{Ki1,Ki2,HLS}.  As we proved in
\cite{GKLU1}, cloaking EM active objects, i.e., objects with generic $J\ne 0$,
 imposes certain hidden boundary conditions on the waves
propagating within
the cloaked region. Any locally finite energy wave  satisfying
Maxwell's equations in the classical, or even weak, sense
must satisfy these
conditions.
We note that in our terminology, the fields $(E,H,D,B)$ are a
finite energy solution if   all the components $E_j$, $D_j$, $H_j$, 
and $B_j$ are locally integrable functions;
the energy of the fields is locally finite; and they satisfy Maxwell's equations in
the classical or weak (distributional) sense. The reason why we concentrate
on such solutions is that the effective medium theory of metamaterial
requires that the scale at which the EM fields change significantly  
is larger than the size of the components (or \emph{cells} ) used implement the 
metamaterial.

For a
cylinder cloaked by what is called the \emph{single coating} construction in
\cite{GKLU1}, and  which corresponds most closely with the cloaking
considered in \cite{PSS1,PSS2,CPSSP,SMJCPSS}, the hidden boundary
conditions are the vanishing of the
angular components
of $E$ and $H$. This is exactly the SHS condition associated with
the angular vector
field $\frac{\partial}{\partial\theta}$.
We show that using a SHS lining has two benefits:
blow up of $B$ of the cloaking surface, which may seriously compromize
effective medium theory for metamaterials, is prevented and secondly 
the farfield pattern of the scattered wave is greatly reduced.
It is shown
in \cite{GKLU1} that there is no theoretical, frequency-dependent
obstruction
to cloaking, but with current technology, cloaking should be
considered as essentially monochromatic, and we will work at fixed frequency $k$.

\section{Single coating of a cylinder}

Let us consider Maxwell's equations on $\R^3$,

\ba
& &\nabla\times   E =ik   B,\\
& &\nabla\times   H =-ik   D ,\\
& &   D =\e     E,\\
& &   B =\mu     H,
\ea
where for simplicity we have taken the conductivity $\sigma=0$.

We consider here EM waves propagating in metamaterials,
which allow one to specify $\e$ and $\mu$ fairly arbitrarily.
These are typically  assembled  from components whose size is somewhat smaller
than the wavelength.
Ideal models of cloaking constructions consist of
prescribed ideal parameter {\newtekst tensors}
   $\e,\mu$, describing
coatings making objects invisible to detection by waves of frequency $k$; physically, these would be
implemented using metamaterials  designed to have $\e,\mu$ as
effective parameters (at  the specified frequency).  Note
that in the cloaking constructions the ideal parameters are singular
on a surface surrounding the object, the \emph{cloaking surface}. 
As a result,  discussed further
below,  we need to consider Maxwell's equations holding
not only in the  \emph{classical} sense
but also in the sense of 
Schwartz distributions \cite{GS}.

In the following,
we describe the non-existence results for  finite energy distributional 
solutions with respect
to the ideal parameter fields, the  consequences of
approximative material configurations, and the role of the SHS lining.  

\subsection{Equations for an ideal single coating }

On $\R^3$, with standard coordinates $x=(x_1,x_2,x_3)$, 
we use cylindrical coordinates $(r,\theta,z)$,
defined by
       $(r,\theta,z)\mapsto
(r\cos \theta,r \sin \theta,z)\in \R^3$.
In \cite{GKLU1} we considered Maxwell's equations on $\R^3\setminus \Sigma$,
\ba
& &\nabla\times \tilde E =ik \tilde B\quad,\quad
\nabla\times \tilde H =-ik \tilde D+\tilde J,\\
& & \tilde D =\tilde \e   \tilde E, \quad
   \tilde B =\tilde \mu   \tilde H,
\ea
where $\tilde \e$ and $\tilde \mu$ correspond to the invisibility
coating materials on the exterior of the infinite cylinder
$N_2=\{r<1\}$
and are Euclidian inside $N_2$. $\tilde \e$ and $\tilde \mu$
are singular at $\Sigma$, namely, for $r =|(x_1,x_2)| \to 1^+$,
\ba
\max\frac{\lambda_j(x)}{\lambda_k(x)} = O\left( (r-1)^{-2}\right)\to \infty,
\ea
where $\lambda_j(x), j=1,2,3,$ are the eigenvalues of $\tilde \e(x)$ or $\tilde
\mu(x)$.
In particular, we considered the question of  when there are fields
$\tilde E,\tilde H,\tilde D,\tilde B$ that together constitute a  finite
energy solution of Maxwell's equations in the sense of distributions.
It was shown in \cite{GKLU1} that, in the presence of internal
currents $\tilde J$ {\newtekst when the cloaked region is, e.g.,\ a ball, such solutions do not
generally exist}.  Let us discuss why  this is so. Even for cloaking
passive objects, i.e., $\tilde J=0$ in {\newtekst the cloaked region}, the singular material
parameters give rise to {\newtekst solutions in Maxwell's equations
that correspond either to surface currents (see below) or to the
blow up in the fields at the cloaking surface. Thus, if the
material does not allow such currents to appear,
 then the resulting fields must not blow up.

Let us next consider in the scattering of a plane wave
by a cloaked cylinder, that is, the case when we have no internal currents
and the EM fields have asymptotics
at infinity corresponding to a sum of a given incident plane} wave
$(\tilde E^{in},\tilde H^{in})$
and scattered wave $(\tilde E^{sc},\tilde H^{sc})$ that satisfies
the Silver-M\"uller radiation condition \cite{ColtonKress}.
It was shown in \cite{GKLU1} that with respect to cylindrical coordinates
$(r,\theta,z)$,
\ba
& &\lim_{r\to 1+} e_\theta(x)\,\cdotp \tilde E(x)=0,\quad
\lim_{r\to 1+} e_\theta(x)\,\cdotp \tilde H(x)=0,
\ea
where $e_\theta$ is the angular unit vector. Let $e_z$
be the vertical unit vector. For general
incoming waves, we have that
\beq
\label{surfacecurrent}
& &\lim_{r\to 1+} e_z(x)\,\cdotp \tilde E(x)-a_e(\frac x{|x|})=0,\\
\nonumber
& &\lim_{r\to 1+} e_z(x)\,\cdotp \tilde H(x)-a_h(\frac x{|x|})=0
\eeq
where $a_e$ and $a_h$ do not vanish.  In the
treatment  of cloaking passive objects \cite{PSS1,PSS2} it
is  assumed {a priori, based on the behavior of rays on the exterior,
that the inside of the cloaked region
is ``dark", that is, the fields
$\tilde E$ and $\tilde H$ vanish
in $\{r<1\}$. } (However, see also \cite{RYNQ,YRQ}, where the behavior
of the fields
within the cloaked region is studied.) Under this assumption, the
$E$ and
$H$ fields  have jumps across $\Sigma$,
\ba
& &b_e=\left(\nu\times \tilde E\right)|_{\Sigma^+}-\left(\nu\times
\tilde E\right)|_{\Sigma^-}
=a_e(x) e_{\theta},\\
& &b_h=\left(\nu\times \tilde H\right)|_{\Sigma^+}-\left(\nu\times
\tilde H\right)|_{\Sigma^-}
=a_h(x) e_{\theta}.
\ea

(Here $\nu$ is the Euclidian normal vector
of $\Sigma$, which is just the radial unit vector $e_r$.)
This implies that
\ba
& &\nabla\times \tilde E =ik \tilde B+\tilde K_{surf}, \quad
\nabla\times \tilde H =-ik \tilde D+\tilde J_{surf},
\ea
in the sense of
distributions on $\R^3$, where the
singular distributions
 $\tilde{K}_{surf}=b_e \delta_{\Sigma},$ $\tilde
J_{surf}=b_h\delta_{\Sigma}$
are terms that can be {\newtekst considered  either as
magnetic and electric 
currents supported
on $\Sigma$,
or, as
below,
idealizing the blow up of $\tilde D$ and $\tilde B$ near $\Sigma$}.
Here, $\delta_\Sigma$ is the distribution defined by
\ba
\int_{\R^3} f(x)\delta_\Sigma \, dx=\int_{\Sigma} f(x)\, dS(x),
\ea
where $dS$ is the Euclidian surface element on the surface
$\Sigma$, for any smooth test function $f$. We refer to such strongly singular 
field components as surface currents.

\subsection{Equations for an  approximate single coating.}

Next,  consider the situation when a metamaterial coating 
only approximates this ideal invisibility coating. We show that the existence of the surfac currents
 for the ideal
cloak causes a blow up  of the fields as the approximation
tends to the singular ideal material $\e,\mu$.

To this end, we modify the construction described in the previous section,
still dealing with a cloaking structure of
   the single
coating type. More precisely, for $1<R<2$, consider an infinite
cylinder in $\R^3$ given,  in  cylindrical coordinates, by
   $N_{2}^R=\{r<R\}$.
On $N_{2}^R$ we choose the  metric to be Euclidian, so that
   the corresponding permittivity and  permeability are homogeneous and
isotropic. In $\R^3 \setminus N_{2}^R$, we take the metric
   $\tilde g$ and the corresponding permittivity
and  permeability
   $\tilde \e$ and $\tilde \mu$  to  be the single coating metric
considered in \cite{GKLU1},
   and the previous section, truncated by being
restricted to $N^R_2$. 
Thus, we start from the materials $\tilde \e$ and $\tilde \mu$
corresponding to the single coating metric $\tilde g$
outside $N_{2}^{R}$ and and replace the metric
with the Euclidian metric in $N_{2}^{R}$.
Then the \emph{anisotropy ratio},
   \ba
   L_R := \sup_{x \in \R^3 \setminus N_{2}^{R}}
   \left(\max\frac{\lambda_j(x)}{\lambda_k(x)} \right) =O\left(
(R-1)^{-2}\right)\to \infty,
   \ea
as the approximate cloaking
construction approaches the ideal, that is, $\radius\to 1^+$.

Next, we consider the wave propagation phenomena that arise  as the approximate cloaking
construction approaches the ideal.

\section{Analysis of solutions}

Assume that $k$ is not a Neumann eigenvalue 
for the Euclidian Laplacian {\newtekst in the 
$2-$dimensional disk $\{r<1\}$;}
 as will be seen later, this is equivalent
   with the condition $(J_0)'(k)\not =0$.
For $1<\radius<2$ fixed, let
\ba
& &N_0=\{r>2\},\\
& &N_{1}^{\radius}=\{\radius<r< 2\},\hbox{ and}\\
& &N_{2}^{\radius}=\{r<\radius\},
\ea
so that the Euclidean space $\R^3$ is the union
$N=\overline N_0\cup \overline N_{1}^{\radius}\cup \overline N_{2}^{\radius}$.
Let $\Sigma_R=\{r=R\}$ be the  (appoximate) cloaking surface and $\nu=\p_r$ be its
Euclidean normal vector on both sides, $\Sigma_R^\pm$. 
To define the approximate cloaking material parameters $\tilde \e^R$ and $\tilde \mu^R$, introduce,
as in \cite{GKLU1},
an auxiliary space $M^R$, and
$\tilde g=\tilde g^\radius$  the Riemannian
metric corresponding to this construction.
$M^R$ is obtained by taking the disjoint union of  three components,
\ba
& &M_0=\{r>2\},\\
& &M_{1}^{\radius}=\{\rho<r < 2\},\\
& &M_{2}^{R}=\{r< \rho\},
\ea
where $\rho=2(\radius-1)$.

The domain $\overline M_0\cup M_{1}^{\radius}$ is a subdomain of $\R^3$,
as is the cylinder $M_{2}^{R}$.
Define, as in \cite{GKLU1}, an abstract manifold $M^R$
by gluing points $(\rho, \theta, z)$ of
the boundary
of $\overline M_0\cup M_{1}^{\radius}$ with points $(\radius, \theta, z)$ of
the boundary
of $M_{2}^{R}$. Equip $M^R$ with the Euclidian metric $g$ and the
corresponding homogeneous, isotropic
permittivity and permeability, $\e, \, \mu$. With respect to the
   cylindrical coordinates, $(r,\theta,z)$, we have
\beq
\label{Number}
& &
g=[g_{jk}]_{j,k=1}^3=\left(\begin{array}{ccc}
1 & 0 & 0\\
0 & r^2& 0 \\
0 & 0 &1\\
\end{array}
\right), \quad
\e=\mu=
\left(\begin{array}{ccc}
r & 0 & 0\\
0 & r^{-1} & 0 \\
0 & 0 & r\\
\end{array}
\right).
\eeq
Next, introduce
a Lipschitz-diffeomorphism $F^R:M^R\to \R^3$, which in
cylindrical coordinates is given by
\ba
& &F^R:M_0\to N_0,\quad F^R|_{M_0}=id,\\
& &F^R:M_{1}^{\radius}\to N_{1}^{\radius},\quad F^R|_{M_{1}^{\radius}}(r,\theta,z)=
(r/2+1\,
     ,\theta,z), \\
& &F^R:M_{2}^{\radius}\to N_{2}^{\radius},\quad F^R|_{M_2^\radius}=id.
\ea
We  define the metric $\tilde g^R$ on $\R^3$ by the formula
$\tilde g^R=(F^R)^*g$, that is,
\ba
\tilde g^{R, jk}(y)=\sum_{p,q=1}^3\frac {\p y^j}{\p x^p}\frac {\p y^k}{\p x^q}\,
g^{pq}(x)
,\quad y=F^R(x),
\ea
where $[\tilde g^{jk}]= [\tilde g_{jk}]^{-1}$ and we use that
$g^{pq}=\delta^{pq}$.
Suppressing for the 
time being the superscript 
$^R$,  the
permittivity and permeability, $ \tilde \e, \,\tilde \mu$ corresponding
to the  metric $\tilde g$
are then given (see, e.g., \cite{GLU1,GLU2}) by
\ba
\tilde \e=\tilde \mu =
|\det(\tilde g_{jk})|^{1/2}\tilde g^{jk}.
\ea
Then the metric $\tilde g$ and permittivity and permeability,
$\tilde \e$,  $\tilde \mu$ are still given by formula
(\ref{Number})
   on  $N_0$ and $N_{2}$.
On $N_{1}$ they are
\ba
& & \tilde g=
\left(\begin{array}{ccc}
4 & 0 & 0\\
0 & 4(r-1)^2 & 0 \\
0 & 0 &  1\\
\end{array}
\right),\\
& &   \tilde \e= \tilde \mu=
\left(\begin{array}{ccc}
   (r-1) & 0 & 0\\
0 & (r-1)^{-1} & 0 \\
0 & 0 &  4(r-1)\\
\end{array}
\right).
\ea
In the following, we consider 
TE-polarized electromagnetic waves.
 This means that, written componentwise
with respect to either coordinate system as
\ba
\tilde E=(\tilde E_1,\tilde E_2,\tilde E_3)=
(\tilde E_r,\tilde E_\theta,\tilde E_z)
\ea
with 
\ba
\tilde E_r=\tilde E_1 \cos(\theta)+\tilde E_2 \sin(\theta),
\quad \tilde E_\theta = r \left(-\tilde E_1 \sin(\theta)+
\tilde E_2 \cos(\theta)  \right), \quad 
\tilde E_z=\tilde E_3,
\ea
the electric field has a nonzero component  only in the $z$-direction,
\ba
\tilde E_1=\tilde E_2=\tilde E_r=\tilde E_\theta=0, \quad
\tilde E_3(x)=\tilde E_3(r, \theta).
\ea
We denote  $\tilde E_3=u$. Then
\ba
\tilde H=\frac 1 {ik}\tilde \mu^{-1} \big(\nabla\times \tilde E\big)
=\frac 1 {ik}\tilde \mu^{-1} \big( e_z \times \nabla u\big).
\ea
We note that $u$ satisfies
the (scalar) Helmholtz equation,
\ba
(\Delta_{\tilde g}+k^2)u=0\quad \hbox{on } \R^3
\ea
where  $\Delta_{\tilde g}$ is the Laplace-Beltrami operator
corresponding to the metric $\tilde g$.

\subsection {Scattering problem}

We consider an incoming TE polarized plane wave. In $N_0$
such a wave has the form
\ba
\tilde E_{in}(r, \theta,z)&=&e^{ikr\cos\theta}=\left(
J_0(kr)+\sum_{n=1}^\infty 2i^nJ_n(kr)\cos(n\theta)\right) e_z,\\
\tilde H_{in}(r, \theta,z)&=&
\frac 1 {ik} \mu^{-1}_0\nabla\times \tilde E_{in},
\ea
or, in terms $\tilde u_{in}= \tilde E_{3, in}$,
\ba
\tilde u_{in}= J_0(kr)+\sum_{n=1}^\infty 2i^nJ_n(kr)\cos(n\theta)
\ea
Here $\mu_0=\e_0=1$.
We look for the solution of the scattering problem,
\beq
\label{tildeM}
& &\nabla\times \tilde E =ik \tilde B,\quad
\nabla\times  \tilde H =-ik  \tilde D,\\ \nonumber
& &  \tilde D =\tilde \e    \tilde E,\quad
 \tilde B =\tilde \mu    \tilde H\,
\eeq
$\hbox{on }\R^3$, where $\tilde \e=\tilde \e^R,\, \tilde \mu=\tilde \mu^R$
so that $\tilde E=\tilde E^R$, etc. Supressing again the index $^R$, 
$\tilde E=\tilde E_{in}+\tilde E_{sc}$,  
$\tilde H=\tilde H_{in}+\tilde H_{sc}$, and
$\tilde E_{sc}$ and $\tilde H_{sc}$ satisfy the Silver-M\"uller radiation condition 
\cite{ColtonKress}.
Analysis of cylindrical cloaking using
Fourier-Bessel series is also in \cite{RYNQ}.

We recall that $\R^3=\overline N_0\cup \overline N_1\cup 
\overline N_2$. 
In the domain $N_0=\{r>2\}$,  one has
\ba
\tilde E_{sc}(r. \theta,z)&=&\left(
\sum_{n=0}^\infty c_nH_n^{(1)}(kr)\cos(n\theta)\right) e_z,\\
\tilde H_{sc}(r,\theta,z)&=&
\frac 1 {ik}\mu_0^{-1} \big(\nabla\times \tilde E_{sc} \big),\\
\tilde u_{sc}&=& 
\sum_{n=0}^\infty c_nH_n^{(1)}(kr)\cos(n\theta)
\ea
Now use  the change of coordinates,
    $F:M\to N$ to define 
the pulled back fields on $M$,
\ba
& &E_{in}=F^*\tilde E_{in},\quad H_{in}=F^*\tilde H_{in},\\
& &E_{sc}=F^*\tilde E_{sc},\quad H_{sc}=F^*\tilde H_{sc},\\
& &E=F^*\tilde E,\quad H=F^*\tilde H.
\ea
In the coordinates $(r', \theta',z')=
F^{-1}(r, \theta,z), \, \theta'=\theta, z'=z$ on $M_0 =F^{-1}(N_0)$,
\ba
& &\nabla\times E =ik  B,\quad
\nabla\times  H =-ik  D ,\\
& &  D =\e_0    E ,\quad
  B =\mu_0    H.
\ea
In $M_1$, i.e. for $r' > \rho$,
\ba
\hspace{-1cm}E(r',\theta',z')&=&\left(J_0(kr')+c_0H_0^{(1)}(kr')+
\sum_{n=0}^\infty
(2i^nJ_n(kr')+c_nH_n^{(1)}(kr'))\cos(n\theta')\right) e_z,\\
\hspace{-1cm}H(r',\theta',z')&=&
\frac 1 {ik}\mu_0^{-1} \big(\nabla\times E \big),\\
\hspace{-1cm}
u(r',\theta')&=& E_3(r',\theta')=
J_0(kr')+c_0H_0^{(1)}(kr')+\\
& &\quad+
\sum_{n=0}^\infty
(2i^nJ_n(kr')+c_nH_n^{(1)}(kr'))\cos(n\theta')
\ea

\ba
\tilde E_r(r,\theta,z)=2 E_r(2(r-1),\theta,z),& & \quad
\tilde E_\theta(r,\theta,z)= E_\theta (2(r-1),\theta,z), \\
& &\tilde E_z(r,\theta,z)= E_z(2(r-1),\theta,z).
\ea
In $N_2$, i.e. for $r < R$,
\beq
\label{Formula}
\tilde E(r,\theta,z)&=&\left(
\sum_{n=0}^\infty a_nJ_n(kr)\cos(n\theta)\right) e_z,\\
\nonumber
\tilde u(r, \theta)&=&
\sum_{n=0}^\infty a_nJ_n(kr)\cos(n\theta),\\
\nonumber
\tilde H(r,\theta,z)&=&
\frac 1 {ik}\mu_0^{-1} \big(\nabla\times \tilde E \big).
\eeq
As $F|_{M_2^R}=id$, the fields $E$ and $H$ and the potential $u$ 
 are also given by (\ref{Formula})  in $M_2$.
 
{\newtekst On $\Sigma^\radius=\p N_2^R$, 
using
the standard transmission conditions for the electric and magnetic fields,
that ensure distributional solutions, are}
\ba
& & \tilde E_\theta|_{\Sigma_\radius+}=\tilde E_\theta|_{\Sigma_\radius-},
\quad \tilde E_z|_{\Sigma_\radius+}=\tilde E_z|_{\Sigma_\radius-};
\\
& &\tilde H_\theta|_{\Sigma_\radius+}= \tilde H_\theta|_{\Sigma_\radius-},
\quad \tilde H_z|_{\Sigma_\radius+}=\tilde H_z|_{\Sigma_\radius-};
\\
& & \tilde D_r|_{\Sigma_\radius+}= \tilde D_r|_{\Sigma_\radius-};\quad
 \tilde B_r|_{\Sigma_\radius+}= \tilde B_r|_{\Sigma_\radius-},
\ea 
we 
get the following transmission conditions for $\tilde u$,
\ba
\tilde u|_{\Sigma_\radius+}= \tilde u|_{\Sigma_\radius-}, \\
4 (R-1) \, \p_r\tilde u|_{\Sigma_\radius+} = R\, \p_r\tilde u|_{\Sigma_\radius-}.
\ea
These correspond to
conditions on $\p M_2^R= \p( \overline M_0 \cup \overline M_1^R)$,
\ba
& &u|_{r=\rho^+}=u|_{r=\radius^-},\\
& &\rho\, \p_ru|_{r=\rho^+}=R\, \p_ru|_{r=\radius^-},
\ea
that give equations for $c_n$ and $a_n$. 

Let us start with $n=0$, which
is of particular interest.
 The above conditions yield for $a_0$ and $c_0$ the equations
\ba
a_0 J_0(kR) &=&J_0(k \rho)+c_0 H_0^{(1)}(k \rho),\\
a_0 Rk ( J_0)'(kR)&=&\rho k (J_0)'(k \rho)+c_0 \rho k (H_0^{(1)})'(k \rho)
\ea
that yield,
{\newtekst when $(J_0)'(k) \not= 0$, that}
\ba
c_0(\radius)=
\frac{\rho (J_0)'(k \rho) J_0(k R) -R J_0(k \rho) (J_0)'(kR)}
{\rho (H_0^{(1)})'(k \rho)J_0(k R) -R H_0^{(1)}(k \rho) (J_0)'(kR)}
=i \pi 
    \frac {1}{\log(k\rho)}(1+o(1)),\\
a_0(R)=
\frac{k \rho  (J_0)'(k \rho) H_0^{(1)}(k \rho) -k\rho  J_0(k \rho) (H_0^{(1)})'(k\rho)}
{\rho (H_0^{(1)})'(k \rho)J_0(k R) -R H_0^{(1)}(k \rho) (J_0)'(kR)}  =
\frac{ 2 \pi}{ (J_0)'(k) \log(k\rho)}(1+o(1)), 
\ea
where we use the asymptotics of Bessel functions near $0$,
see \cite[pp.360--361]{AS}. Here, $o(1)$ means that the quantity
goes to zero as $\radius\to 1^+, \, \rho \to 0^+,$. Similarly, 
 $a_n$ and $c_n$ satisfy the equations,
\ba
  a_n J_n(kR) &=&J_n(k \rho)+c_n H_n^{(1)}(k \rho),\\
a_n Rk ( J_n)'(kR)&=&\rho k (J_n)'(k \rho)+c_n \rho k (H_n^{(1)})'(k \rho),
\ea
that yield, {\newtekst for the generic $k$,}
 that 
\beq
\label{asymptoticsn}
c_n(\radius)=
\frac{\rho (J_n)'(k \rho) J_n(k R) -R J_n(k \rho) (J_n)'(kR)}
{\rho (H_n^{(1)})'(k \rho)J_n(k R) -R H_n^{(1)}(k \rho) (J_n)'(kR)}
=O(\rho^{2n}),\\
\nonumber
a_n(R)=
\frac{k \rho  (J_n)'(k \rho) H_n^{(1)}(k \rho) -k\rho  J_n(k \rho) (H_n^{(1)})'(k\rho)}
{\rho (H_n^{(1)})'(k \rho)J_n(k R) -R H_n^{(1)}(k \rho) (J_n)'(kR)}  = O(\rho ^n).
\eeq
This implies
that the scattered fields (far-field patterns)  
$\tilde E_{sc}, \tilde H_{sc}$ in $\overline{N_0} \cup N_1^R$ 
and the 
transmitted fields $\tilde E, \tilde H$ in $N_2^R$,
which, as we recall, depend on $R$,
go to zero as the
approximate cloaking construction
tends to the ideal material parameters, i.e. $R \to 1^+$.
A similar result was obtained in \cite{RYNQ}.

Next, we consider the behavior of the fields 
$\tilde E^R, \tilde H^R, \tilde D^R, \tilde B^R$
near $\Sigma_\radius=\p N_2^R$. Supressing again the superscript
$^R$, 
we write the electric and magnetic fields as
\beq
\label{19.6}
\tilde E(r,\theta,z)&=&\sum_{n=0}^\infty \tilde  E^n(r,\theta,z),\quad\hbox{where} \quad
\tilde E^n(r,\theta,z)=f_n(kr)\cos(n\theta) e_z;\\
\nonumber
\tilde H(r,\theta,z)&=&\sum_{n=0}^\infty \tilde H^n(r,\theta,z),\quad\hbox{where} \quad
\tilde H^n(r,\theta,z)=\frac 1 {ik}\tilde \mu^{-1}\left(\nabla\times \tilde E^n(r,\theta,z) \right),
\eeq
 with similar notations
for the scattered and incoming fields,
$\tilde E_{sc}^n$, $\tilde H_{sc}^n$, etc.
{On $M$, the decomposition (\ref{19.6}) gives rise to a similar decomposition
of $E$ and $H$, which we analyze for each value of $n$.
First, we consider  the terms corresponding to $n=0$.
On $\overline M_0 \cup \overline M_1^\radius$, at $y=F_1^{-1}(x), y=(r', \theta', z'),\,
x \in \overline N_0 \cup \overline N_1^R$,
\ba
E^0_{in, z}(y)&=&J_0(k r')=O(1),
\\
E^0_{sc, z}(y)&=&c_0(R) H^{(1)}_0(k r')= - \frac{\ln{k r'}}{\ln{k \rho}}\left(1+o(1)\right).
\ea
Observe that, since $r' \geq \rho$, $E^0_{sc, z}(y)$ is uniformly bounded for $R \to 1^+$.
With the magnetic field $H^0$ having a non-zero component only in $\theta$, one has
\ba
H^0_{in, \theta}(y)&=& i r' \left(J_0\right)'(k r') =O((r')^2);\\
H^0_{sc, \theta}(y)&=&  i r' c_0(R) \left(H^{(1)}_0\right)'(k r')= 
\frac{i}{k \ln{(k \rho)}}
\left(1+o(1)\right).
\ea
On $M_2^R$,
\ba
E^0_z(y)&=& a_0(R) J_0(k r')=\frac{O(1)}{\ln{(k \rho)}}; \\
H^0_\theta(y)&=& i a_0(R) r' \left(J_0\right)'(k r') =\frac{O(1)}{\ln{(k \rho)}}.
\ea
Returning to $N$ and again using the transformation rules for $E$ and $H$, we see that
$\tilde E^0,\, \tilde H^0$ are uniformly bounded, with respect to $R$, in 
$\overline N_1^R \cup \overline N_2^R$. 
}

Now consider the magnetic flux density, $\tilde B=\tilde \mu \tilde H$ which
 has a similar decomposition. In particular, on $N_1^R$, one has
\beq
\label{19.6.1}
& &\tilde B_{in, \theta}^0(r,\theta)=
\tilde \mu \tilde H_{in, \theta}^0(r,\theta)=
\tilde \mu H^0_{in, \theta}\left(2(r-1), \theta \right) 
=O(r-1), \\ \nonumber
& &\tilde B_{sc, \theta}^0(r,\theta)=
\tilde \mu \tilde H_{sc, \theta}^0(r,\theta)
=\tilde \mu H^0_{sc, \theta}\left(2(r-1), \theta \right) =
\frac{i}{k (r-1) \ln{k \rho}}(1+o(1)).
\eeq
{Pointwise, on $N_2^R$,
\ba
\tilde B^0_\theta(r, \theta) =\frac{O(r)}{\ln{(k \rho)}},
\ea
tending to $0$ when $R \to 1^+$. However, to see how $\tilde B_{\theta}^0$ behaves
as a distribution  as $\radius \to 1^+$,
observe that (\ref{19.6.1}) implies that
$
\int_{1/2}^{3/2} \tilde B_{\theta}^0(r, \theta)\, d r 
$
is uniformly bounded as $R \to 1^+$,}
while for any $0<\kappa<\frac 12$,
\ba
\int_{1-\kappa}^{1+\kappa} \tilde B_{\theta}^0(r, \theta)\, d r 
= \frac ik \int_{\rho/2}^{\kappa} \frac 1{(\log \rho)t} dt +o(1) \to
\frac ik \quad \hbox{when }\radius=\rho/2+1\to 1^+.
\ea
This implies that
\ba
\lim_{\radius\to 1^+}\tilde B^0_\theta=\frac ik\delta_\Sigma+
\tilde B^0_{b, \theta},
\ea
in the sense of distributions,
where $\delta_\Sigma$ is the delta-function of the cylinder $\Sigma=\{r=1\}$
and
$\tilde B^0_{b, \theta}$ is a bounded function.

At last, consider $\tilde D^0$ which has only the
$z-$component
different from $0$. In $N_1^R$,
\ba
\tilde D^0_{in, z}(r, \theta)= \tilde \e \tilde E^0_{in,  z}(r, \theta)=
(r-1) E^0_{in, z}(2(r-1), \theta)=O(r-1);\\
\tilde D^0_{sc, z}(r, \theta)= \tilde \e \tilde E^0_{sc,  z}(r, \theta)=
(r-1) E^0_{sc, z}(2(r-1), \theta)=
\frac{O((r-1) \ln{(r-1)})}{\ln{(k \rho)}},
\ea
while in $N_2^R$,
\ba
\tilde D^0_{z}(r, \theta)= \tilde \e \tilde E^0_z(r, \theta)=
r E^0_z(r, \theta)=\frac{O(1)}{\ln{(k \rho)}}.
\ea
Thus,
when $R \to 1^+$,  $\tilde D^0_{in, z}$ has a uniform limit in
$N_0 \cup N_1$,
 and $\tilde D^0_{sc, z}, \,\tilde D^0_{z}$ 
uniformly tend to $0$ in $N_0 \cup N_1,\, N_2$, respectively.

For $n\geq 1$, using (\ref{asymptoticsn}), we 
obtain the following asymptotics for $\tilde E$, etc in various
subdomains of $N$:

In $N_1^R$, where $r>R,$ i.e. $2(r-1) > \rho$,
\ba
\tilde E^n_{in, z}=O((r-1)^n), \quad 
\tilde E^n_{sc, z}=O\left(\frac{\rho^{2n}}{(r-1)^n}\right);\\
\tilde H^n_{in, r}=O((r-1)^{n-1}), \quad 
\tilde H^n_{sc, r}=O\left(\frac{\rho^{2n}}{(r-1)^{n+1}}\right),\\
\tilde H^n_{in, \theta}=O((r-1)^n), \quad
\tilde H^n_{sc, \theta}=O\left(\frac{\rho^{2n}}{(r-1)^n}\right);\\
\tilde D^n_{in, z}=O((r-1)^{n+1}), \quad
\tilde D^n_{sc, z}=O\left(\frac{\rho^{2n}}{(r-1)^{n-1}}\right);\\
\tilde B^n_{in, r}=O((r-1)^n), \quad
\tilde B^n_{sc, r}=O\left(\frac{\rho^{2n}}{(r-1)^n}\right),\\
\tilde B^n_{in, \theta}=O((r-1)^{n-1}), \quad
\tilde B^n_{sc, \theta}=O\left(\frac{\rho^{2n}}{(r-1)^{n+1}}\right)
\ea
As for $N_2^R$, we have
\ba
\tilde E^n_{z}=O(\rho^n);\quad \tilde D^n_{z}=O(\rho^n);\\
\tilde H^n_{r}=O(\rho^n),\quad \tilde H^n_{\theta}=O(\rho^n);\\
\tilde B^n_{r}=O(\rho^n),\quad \tilde B^n_{\theta}=O(\rho^n)
\ea
These formulae imply that there is a uniform limit of $\tilde E^n,\,
\tilde H^n,\,\tilde D^n,\,\tilde B^n$ when $R \to 1^+$ and, moreover,
the scattered fields in $\overline{N_0 \cup N_1}^{int}$ and transmitted fields in 
$N_2$ tend to $0$.
 These formulae also imply that the series
\ba
\sum_{n=1}^\infty \tilde E^{n}_{sc},\quad
\sum_{n=1}^\infty \tilde H^{n}_{sc},\quad
\sum_{n=1}^\infty \tilde B^{n}_{sc},\quad
\sum_{n=1}^\infty \tilde D^{n}_{sc},
\ea
in $\overline{N_0 \cup N_1}$; and 
\ba
\sum_{n=1}^\infty \tilde E^{n},\quad
\sum_{n=1}^\infty \tilde H^{n},\quad
\sum_{n=1}^\infty \tilde B^{n},\quad
\sum_{n=1}^\infty \tilde D^{n},
\ea
in $N_2$,
all converge to zero, as $\radius\to 1^+$, in the sense of distributions, i.e., in
${\cal D}'(N,dx)$.

Summarizing, we see that, in the sense of distributions,
\ba
& &\lim_{\radius\to 1^+}  \tilde E^\radius= \tilde  E_{b},\quad
\lim_{\radius\to 1^+}\tilde   H^\radius= \tilde  H_{b},\\
& &\lim_{\radius\to 1^+} \tilde  D^\radius= \tilde  D_{b}
  - \frac 1{ik}  \tilde   J_{surf},
\quad \tilde   J_{surf}=0,\\
& &\lim_{\radius\to 1^+} \tilde  B^\radius= \tilde  B_{b}+\frac 1{ik}{\tilde   K_{surf}}
,\quad
\tilde K_{surf}= -\delta_\Sigma.
\ea
Here $ \tilde E_{b}$, $\tilde H_{b}$,
$ \tilde D_{b}$, and $ \tilde B_{b}$ coincide with 
$\tilde E_{in}, \tilde H_{in}, \tilde B_{in},$ and $\tilde B_{in}$, correspondingly, in $\overline{N_0 \cup N_1}$ and are equal to $0$
in $N_2$. Thus, in particular, they satisfy equations (\ref{tildeM})
separately in $N_0 \cup N_1$ and $N_2$.

Note that, sending an TM-polarized wave, we get that 
$\tilde   J_{surf}=-\delta_\Sigma,\, \tilde   K_{surf}=0$.
Moreover, for a general incoming electromagnetic wave,
the corresponding solutions $\tilde E^R, \tilde H^R, \tilde D^R, \tilde B^R$
tend  in $L^1(\R^3)$ to $\tilde E_{lim}, \tilde H_{lim}, 
\tilde D_{lim}, \tilde B_{lim}$ which satisfy
\ba
& &\nabla\times \tilde E_{lim} =ik\tilde   B_{lim}+\tilde K_{surf} ,\quad
\nabla\times  \tilde H_{lim} =-ik \tilde  D_{lim}
+\tilde J_{surf},\\
& & \tilde  D_{lim} =\tilde \e  \tilde   E_{lim} ,\quad
 \tilde  B_{lim} =\tilde \mu    \tilde H_{lim},
\ea
with $\tilde J_{surf}= b_e \delta_\Sigma,\, \tilde K_{surf}= b_h  \delta_\Sigma$.

\section{Numerical results}

We next use
the analytic expressions found above to compute the fields 
when a plane wave, with vertically polarized E-field,
$E_{in}(r,\theta,z)=e^{ikr\cos \theta}\vec e_z$ having 
wavenumber $k=3$,
is incident 
to a cylinder $\{r<\radius\}$ that is coated with an approximative invisibility
cloaking layer located in $\{\radius<r<2\}$.
We  then numerically simulate the cases where $\radius=1.01$ and
   $\radius=1.05$. In the simulations we have used
Fourier series representation to  order $6$, that is, the
fields are represented using 
{trigonometric polynomials of degree less than or equal to six, 
$\sum_{|n|\le 6}f_n(r)e^{in\theta}$.}
In the tables below, we give the real parts of the
$\theta$-component
of the total fields and the scattered $B$-field
on the line $\{(x,0,0):\ x\in [0,3]\}$,
{first in the absence of a physical layer inside the
metamaterial and then when  an SHS lining is included.
We note that in the case of the SHS lining, the fields
are as was  claimed of \cite{PSS1,CPSSP,SMJCPSS,W} without reference
 to a lining, namely
 zero inside the cylinder $\{r<\radius\}$.
In Figs. 1,2,  we see clearly the development of a delta-type
distribution on the interface when we do not have the
SHS lining and  the approximative
cloaking approaches the ideal, i.e., $\radius \to 1^+$.}
Also, we see that far away from the coated cylinder
in both cases the scattered field goes to zero,
{\newtekst but much more quickly when the SHS lining is used.}

\vfil\eject
Below we give the numerically computed Fourier coefficients
of the scattered waves.
When $L=1600$, that is,
   $\radius=1.05$, we have:

\medskip
\begin{tabular}{crr}
$n$ &    $c_n$ with SHS lining &   $c_n$ with no SHS lining \\
\hline
0 & $  -0.0042 - 0.0644i$ & $-0.6308 - 0.4826i$\\
1 & $  -0.0099 + 0.1407i$ & $-0.0067 - 0.1154i$\\
2 & $  -0.0016 + 0.0000i$ & $-0.0008 - 0.0000i$\\
3 & $   0.0000 + 0.0000i$ & $ 0.0000 + 0.0000i$\\
4 & $   0.0000 + 0.0000i$ & $ 0.0000 + 0.0000i$\\
5 & $   0.0000 + 0.0000i$ & $ 0.0000 + 0.0000i$\\
6 & $   0.0000 + 0.0000i$ & $ 0.0000 + 0.0000i$\\
\hline\\
$(\sum |c_n|^2)^{1/2}$  & $ 0.1551$ &  0.8026\\
\end{tabular}

\bigskip
When $L=40,000$, that is,
   $\radius=1.01$, we have:

\medskip

\begin{tabular}{crr}
$n$ &    $c_n$ with SHS lining &   $c_n$ with no SHS lining \\
\hline
0 & $    -0.0000 - 0.0028i$ & $  -0.2318 - 0.4220i$\\
1 & $    -0.0000 + 0.0057i$ & $  -0.0000 - 0.0049i$\\
2 & $     0.0000 + 0.0000i$ & $ 0.0000 + 0.0000i$\\
3 & $     0.0000 + 0.0000i$ & $ 0.0000 + 0.0000i$\\
4 & $     0.0000 + 0.0000i$ & $ 0.0000 + 0.0000i$\\
5 & $     0.0000 + 0.0000i$ & $ 0.0000 + 0.0000i$\\
6 & $     0.0000 + 0.0000i$ & $ 0.0000 + 0.0000i$\\
\hline\\
$(\sum |c_n|^2)^{1/2}$  & $ 0.0063 $ &   0.4815\\
\end{tabular}

\bigskip
{The results show that for $\radius$  close to 1,
including the SHS  lining strongly reduces the farfield of the scattered
wave; the approximative
invisibility cloaking functions much better with such a lining than without, 
even for cloaking passive objects.}

\begin{figure}[htbp]
\begin{center}
\centerline{\includegraphics[width=.55\linewidth]{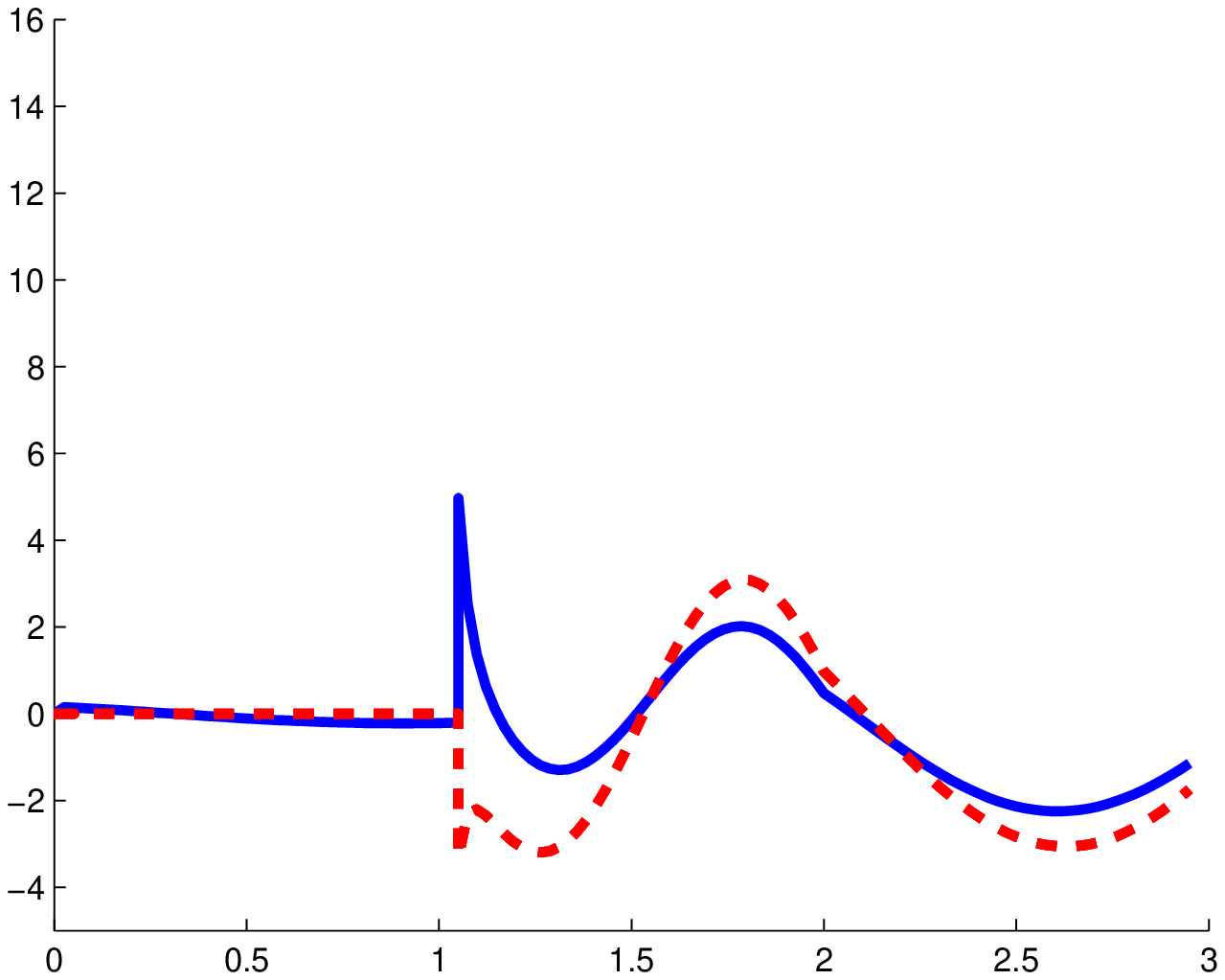}
\includegraphics[width=.55\linewidth]{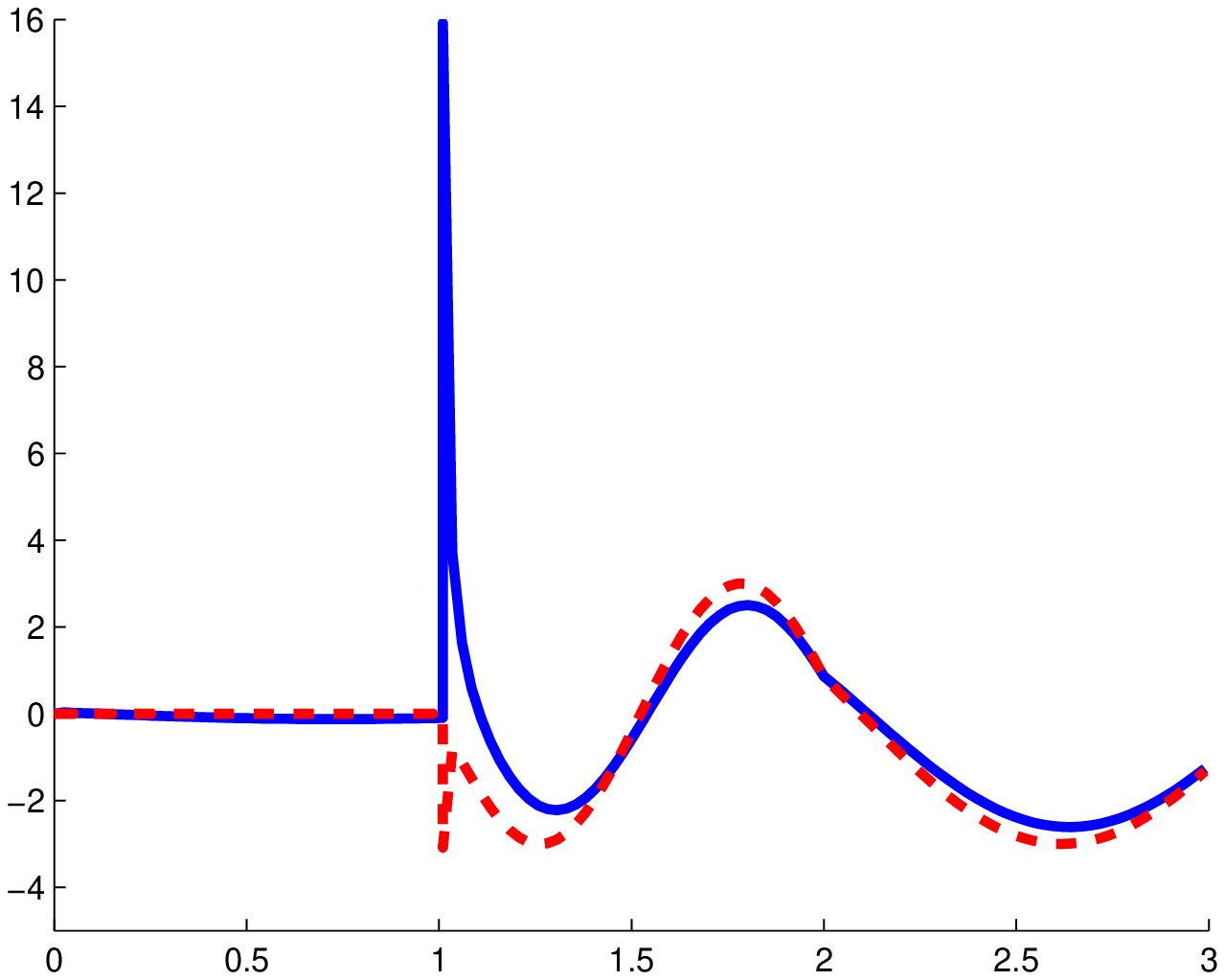}}
%\centerline{\includegraphics[width=.55\linewidth]{To_paper_total_fields_L_small.pdf}
%\includegraphics[width=.55\linewidth]{To_paper_total_fields.pdf}}
\end{center}
\vspace{-1cm}

\caption{The $\theta$-component
of the total $B$-field on the line $\{(x,0,0):\ x\in [0,3]\}$.
Blue solid line correspond the field with no physical lining
at $\{r=\radius\}$. Red dashed line correspond the field with
Soft-and-Hard lining on  $\{r=\radius\}$. At the left figure, the maximal
anisotropy ratio is $L=1600$ and $\radius=1.05$.
At the right figure, the maximal
anisotropy ratio is
   $L=40,000$ and $\radius=1.01$.}
\end{figure}

\begin{figure}[htbp]
\begin{center}
\centerline{\includegraphics[width=.55\linewidth]{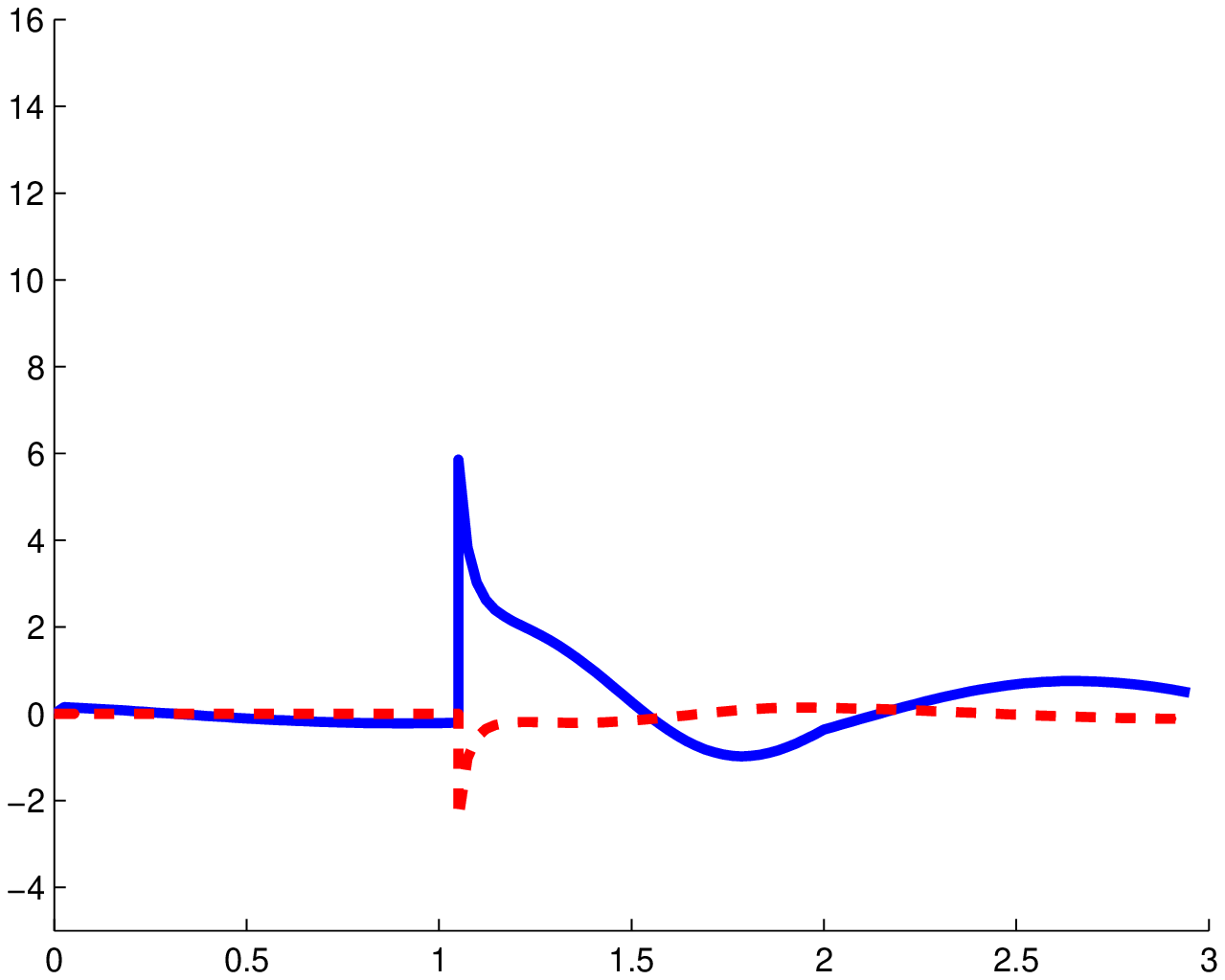}
\includegraphics[width=.55\linewidth]{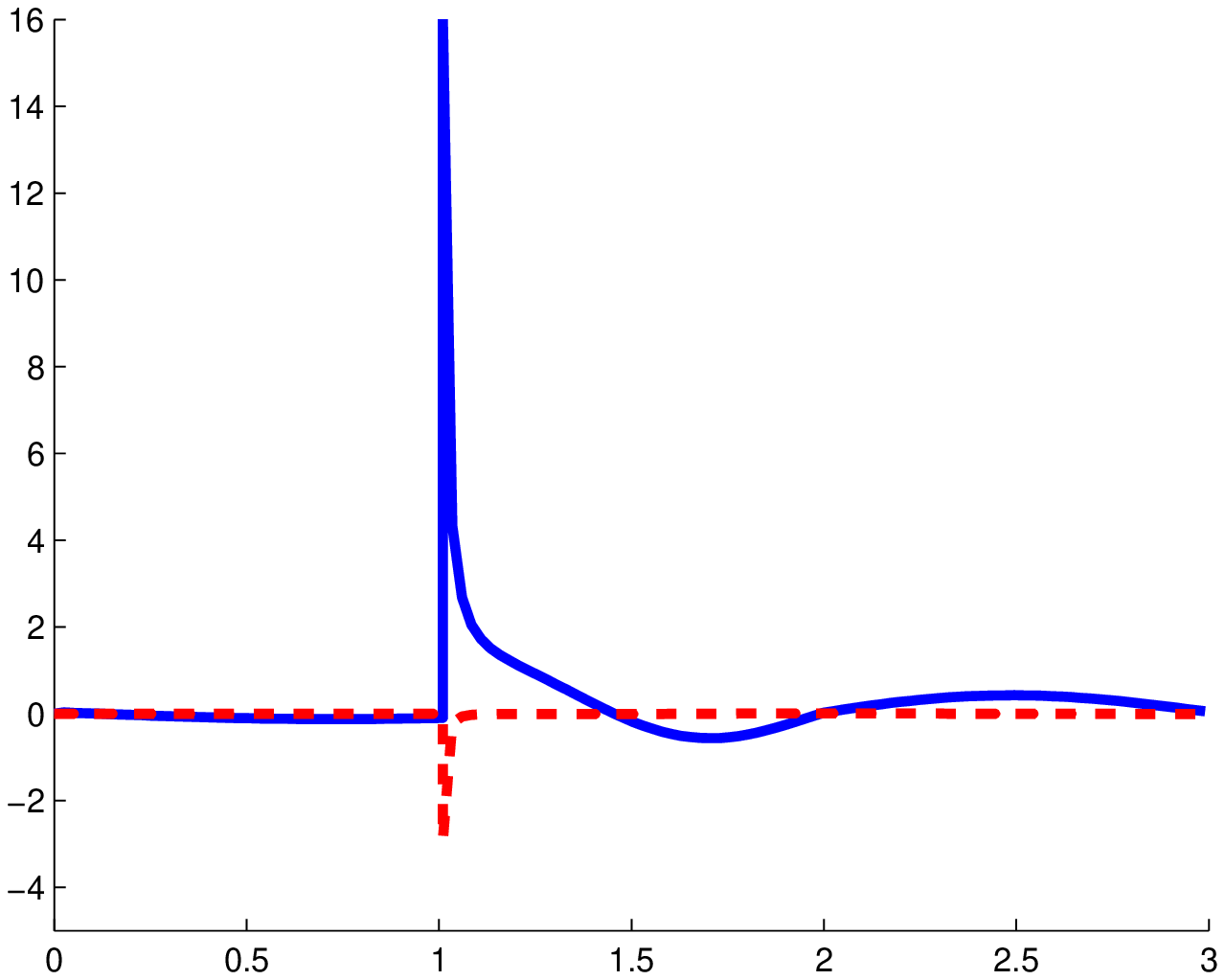}}
%\centerline{\includegraphics[width=.55\linewidth]{To_paper_scattered_fields_L_small.pdf}
%\includegraphics[width=.55\linewidth]{To_paper_scattered_fields.pdf}}
\end{center}
\vspace{-1cm}

\caption{The $\theta$-component
of the scattered $B$-field on the line $\{(x,0,0):\ x\in [0,3]\}$.
Blue solid line correspond the field with no physical lining
at $\{r=\radius\}$. Red dashed line correspond the field with
Soft-and-Hard lining on  $\{r=\radius\}$. At the left figure, the maximal
anisotropy ratio is $L=1600$ and $\radius=1.05$.
At the right figure, the maximal
anisotropy ratio is  $L=40,000$ and $\radius=1.01$.
}
\end{figure}

\section{Discussion}

\subsection{Comparison of results with and without SHS.}

One  observes that,  without the  SHS  lining,  the $B$-field grows as the
approximate single coating  tends
more closely to the ideal
invisibility cloak, i.e.,
as  the anisotropy ratio $L$ becomes larger.
 In both Figs. 1 and 2, the peak near $r=1$
without the SHS  lining shows quite clearly how the
delta-distribution
in the $B$-field develops.

Note that  the value of the anisotropy ratio $L$ is quite large in our
simulation,  but the resulting
fields are still not extremely large.
So, it is not surprising that the non-existence results in
\cite{GKLU1},  predicting
the blow up of the fields, were not observed in the experiment
\cite{SMJCPSS}.
However, it seems likely to become  more significant
as  cloaking
technology develops. Also, the SHS boundary
lining has the additional benefit  in our simulations of making the  scattered
wave smaller outside of the metamaterial construction.
Indeed, the scattered field when using the
SHS  boundary lining
is less than 2\% of the scattered field without the lining.
Thus, implementation of a lining significantly improves the cloaking effect.

\noindent
\subsection{Significance of the surface currents $\tilde J_{surf}$ and $\tilde K_{surf}$}
As $\radius\to 1^+$, for generic
     incoming waves the magnetic and electric flux densities
converge to fields that
{contain delta-function  type distributional components
supported on the surface $\Sigma$.}
We phrase this by saying that  \emph{ surface currents}  appear.
{\newtekst  
If
the metamaterial construction allows this, 
then we interpret this literally. This holds,
e.g.,  if the metamaterials used have components
near $\Sigma$ that approximate a SHS surface, such as strips of PEC
and PMC materials.}
Alternatively, {if no such currents can
appear in the material,} the $\tilde D$ and $\tilde B$ fields will blow up
as the approximation of the coating material goes to the limit
$\radius\to 1^+$.

Effective medium theory for composite materials is
 proven only when the limiting fields are relatively
smooth  \cite{M-book}.
{\newtekst Such rigorous effective medium theory has not yet been established for metamaterials, but the limited
work so far, e.g.,  \cite{SP}, clearly indicate that this same restriction will hold there as well.
 One can then  interpret the blow up of
fields
as a challenge to the validity of the material parameters that have been ascribed to the metamaterilas currently employed.
Indeed, fields having a blow up
are very rapidly changing functions  near the cloaking surface.
Thus making a physical cloaking construction that would operate
well with such fields would require metamaterials whose cell size
becomes very small close to  the cloaking surface.}

{\newtekst The simplest way to avoid these issues would be to
include the SHS lining when constructing the cloaking device.}

\bigskip

\subsection{Summary}
We have considered two cases when cloaking an infinite cylinder:

\bigskip
\noindent {\bf (1)} An infinite cylinder of air  or vacuum, is coated with metamaterial in $\{R<r<2\}$ 
but has no lining on the  interior surface of the metamaterial coating. 
In the limit $R\to 1^+$, solutions to Maxwell's equations
   have singular current terms $K_{surf}$ and $J_{surf}$ that represent either
surface currents or the blow up of the $D$ and $B$ fields. 
A standard assumption in homogenization theory is that the length scale, $d$,
of the substructures (or cells) from which a composite medium is formed, is much less than the
free space wavelentgnt $\lambda$ of the EM field \cite{M-book}. In treatments
of homogenization for metamaterials, e.g., \cite{SP}, it has been observed that effective
material parameters can often be obtained even when $d$ is not greatly less than $\lambda$.
Although not explicitly stated, it is  required that sampled surface intergrals
of $E,H,B$, and $D$ not vary greatly from point to point within a metamaterial cell. The
blow up of $B$ that we have shown occurs when cloaking wihout an SHS lining thus presents
a challenge to the effective medium intepretation of the metamaterials employed.

\medskip

\noindent {\bf (2)}  An infinite cylinder of air or vacuum is coated with metamaterial in $\{R<r<2\}$ 
and a SHS-lining on the  interior of the cloaking surface. The lining can be considered as parallel
   PEC and PMC strips, that allow surface currents in the $z$-directions. In
   this case, when $R\to  1^+$, the total $E$ and $H$ fields at the boundary
   have very  small $\theta$-components, that is, in the limit the
   tangent components of $E$ and $H$ are $z$-directional. The non-zero
   tangential boundary values of $E$ and $H$ correspond physically to surface
   currents, that are now allowed because of the SHS lining. Since the
   surface lining and fields are now compatible, the fields do not blow
   up. In addition, the amplitude of the farfield pattern is greatly reduced.

\bibliographystyle{amsalpha}

\vskip.2in

\noindent{\sc Department of Mathematics}

\noindent{\sc University of Rochester}

\noindent{\sc Rochester, NY 14627, USA

\noindent\emph{Email:}\tt{allan@math.rochester.edu}}

\vskip.2in

\noindent{\sc Department of Mathematical Sciences}

\noindent{\sc University of Loughborough}

\noindent{\sc Loughborough, LE11 3TU, UK

\noindent\emph{Email:}\tt{Y.V.Kurylev@lboro.ac.uk}}

\vskip.2in

\noindent{\sc Institute of Mathematics}

\noindent{\sc Helsinki University of Technology}

\noindent{\sc Espoo, FIN-02015, Finland

\noindent\emph{Email:}\tt{Matti.Lassas@tkk.fi}}

\vskip.2in

\noindent{\sc Department of Mathematics}

\noindent{\sc University of Washington}

\noindent{\sc Seattle, WA 98195, USA

\noindent\emph{Email:}\tt{gunther@math.washington.edu}}

\vskip.2in
\noindent\emph{Acknowledgements:} A.G. was supported by NSF-DMS;
M.L. by CoE-program 213476 of the Academy of Finland; and G.U. by NSF-DMS and a Walker Family Endowed Professorship.

\end{document}